\documentclass[11pt]{article}

\usepackage{latexsym}
\usepackage{array}
\usepackage{amsmath,amssymb,amsthm,amsfonts}        
\usepackage{vmargin}
\usepackage{fancyhdr}
\usepackage{accents}
\usepackage{ifthen}                                 
\usepackage{xspace}                                 
\setpapersize[portrait]{USletter}
\newcommand{\thmnumbering}{section}
\renewcommand{\thmnumbering}{}
    \setmarginsrb{1in}{1in}{1in}{1in}{0in}{0in}{12pt}{24pt}
\newcommand{\Claim}[1]{Claim~\ref{clm:#1}}
\newcommand{\ClaimName}[1]{\label{clm:#1}}
\newcommand{\Equation}[1]{Eq.\:\eqref{eq:#1}}
\newcommand{\EquationName}[1]{\label{eq:#1}}

    \newcommand{\thmbelow}{6pt}
\newcommand{\thmabove}{7.5pt}
    
    \makeatletter    
    \renewcommand{\maketitle}{
        \begin{center}
        \begin{minipage}[t]{6.5in}
        \begin{center}
        \vspace*{15pt}
        {\@title \par}
        \vspace*{20pt}
        {\@author}
        \vspace*{3pt}
        \end{center}
        \end{minipage}
        \end{center}
    }
    \makeatother

    \newtheoremstyle{mythmstyle}
      {\thmabove}   
      {\thmbelow}   
      {}            
      {}            
      {\bfseries}   
      {. }          
      {2.5pt}       
      {\thmname{#1}\thmnumber{ #2}\thmnote{ \normalfont (#3)}}   
    \theoremstyle{mythmstyle}
    \ifthenelse{\equal{\thmnumbering}{section}}
        {\newtheorem{theorem}{Theorem}[section]\numberwithin{equation}{section}}
        {\newtheorem{theorem}{Theorem}}
    
    \newtheorem{claim}[theorem]{Claim}
    
\newcommand{\bR}{\mathbb{R}}
\newcommand{\set}[1]{\left \{ #1 \right \}}                     
\newcommand{\setst}[2]{\left\{\; #1 \,:\, #2 \;\right\}}        
\newcommand{\abs}[1]{\lvert #1 \rvert}
\newcommand{\card}[1]{\abs{#1}}
\newcommand{\transpose}{^{\textsf{T}}}                                    
\newcommand{\defeq}{\,:=\,}                                     
\renewcommand{\th}{\ifmmode{^{\textrm{th}}}\else{\textsuperscript{th}\ }\fi}
\newcommand{\bZ}{\mathbb{Z}}

\newenvironment{LPmin}[1]{\begin{LP}{min}{#1}}{\end{LP}}
\newenvironment{LP}[2]
{\setlength{\extrarowheight}{1mm}\begin{equation*}
\begin{array}{l >{\displaystyle}l @{\quad} >{\displaystyle}l @{\qquad} l}
\text{#1}  & \multicolumn{3}{l}{\displaystyle #2} \\[0mm]
\text{s.t.} }
{\end{array}\end{equation*}}

\title{\Large  A Randomized Rounding Algorithm for the\\ Asymmetric Traveling Salesman Problem}
\author{Michel X. Goemans \qquad Nicholas J. A. Harvey \qquad Kamal Jain \qquad Mohit Singh}
\date{}

\begin{document}
\maketitle

\begin{abstract}
We present an algorithm for the asymmetric traveling salesman problem
on instances which satisfy the triangle inequality.
Like several existing algorithms, it achieves approximation ratio $O(\log n)$.
Unlike previous algorithms, it uses randomized rounding.
\end{abstract}

\section{Introduction}
Let $V$ be a set of $n$ vertices and let
$c : V \times V \rightarrow \bR_+$ be a cost function.
We assume the triangle inequality:
$c_{i,j} \leq c_{i,k} + c_{k,j}$ for all vertices $i,j,k$.
The asymmetric traveling salesman problem (ATSP) is to solve
$\min_\pi \sum_{v \in V} c_{v,\pi(v)}$
over all cyclic permutations $\pi$ on $V$.
A subgraph $\setst{ (v,\pi(v)) }{ v \in V }$ is called a tour;
we seek a minimum cost tour.

We will use the following standard notation.
For $U \subseteq V$, define
\begin{align*}
\delta^+(U) &~=~ \setst{ (v,w) }{ v\in U ,\: w \not \in U }, \\
\delta^-(U) &~=~ \setst{ (w,v) }{ v\in U ,\: w \not \in U }.
\end{align*}
For any vector $x \in \bR_+^{V \times V}$ and $F \subseteq V \times V$,
we use the notation $x(F) = \sum_{e \in F} x_e$.

The Held-Karp linear programming relaxation of ATSP is as follows.
\begin{LPmin}{c \transpose x}
&x(\delta^-(\set{v})) &= &x(\delta^+(\set{v})) \quad \forall v \in V \\
&x(\delta^+(U)) &\geq &1 ~\quad\qquad\qquad\forall\, \emptyset \neq U \subsetneq V \\
&x &\geq &0
\end{LPmin}
By standard shortcutting arguments, we may assume that $x_{(v,w)} \leq 1$ for all $v,w$,
and that $x(\delta^+(\set{v}))=1$ for all $v$.

Several polynomial-time algorithms
\cite{Frieze} \cite[pp.~125]{Williamson} \cite{Kaplan} \cite{FeigeSingh}
are known for computing a tour
whose cost is at most a factor $O(\log n)$ larger than the optimum.
In addition, several proofs
\cite{Wolsey,Shmoys}
are known showing that the integrality gap
of the Held-Karp relaxation is $O(\log n)$.
This note provides another such algorithm and another such proof of the integrality gap.

\section{The Algorithm}
The algorithm proceeds in two steps. In the first step, we round the fractional solution
using a simple randomized rounding schema to obtain \emph{nearly-balanced} graph. In the second step,
we solve the \emph{patch up} problem to make the graph Eulerian. The algorithm succeeds in returning 
a connected Eulerian subgraph of small cost with high probability.

\subsection{Constructing a nearly balanced graph}
Let $x$ be any feasible solution to this linear program.
Since $x$ is balanced at every vertex, this implies that $x$ is Eulerian,
i.e., $x(\delta^+(U))=x(\delta^-(U))$ for all $U \subseteq V$.
So $x$ is a fractional solution for which all cuts are perfectly balanced.
We now use $x$ to construct an integral solution $z$ for which all cuts are nearly-balanced, i.e.,
for each cut $U \subsetneq V$, $\frac{z(\delta^{+}(U))}{z(\delta^-(U))}\leq 2$.
Moreover, the cost of $z$ is at most $O( \log n ) \cdot c \transpose x$.

The first observation is that $x$ has equivalent cut values to an undirected graph.
Formally, for $U \subseteq V$, define
$\delta(U) = \setst{ \set{v,w} }{ v\in U ,\: w \not \in U }$.
For $y \in \bR_+^{\binom{V}{2}}$ and $F \subseteq \binom{V}{2}$,
let $y(F) = \sum_{e \in F} y_e$.

\begin{claim}
\ClaimName{equiv}
Since $x$ is Eulerian, there exists $y \in \bR_+^{\binom{V}{2}}$ such that
$y(\delta(U)) = x(\delta^+(U))$ for all $U \subseteq V$.
\end{claim}
\begin{proof}
Define $y_{\set{v,w}} = (x_{v,w}+x_{w,v})/2$ for all $v,w$.
Then
$$
y(\delta(U))
 ~=~ \sum_{v \in U ,\: w \not \in U} \frac{x_{v,w} +x_{w,v}}{2}
 ~=~ \frac{1}{2} \Big( x(\delta^+(U)) + x(\delta^-(U)) \Big)
 ~=~ x(\delta^+(U)),
$$
as required.
\end{proof}

We now apply a random sampling result of Karger \cite{Karger}.
For convenience, we reprove it here in our notation.
For any undirected graph with minimum cut value $c$,
Karger \cite{Karger} shows that the number of cuts
of value at most $\alpha c$ is less than $\binom{n}{2\alpha}$.
This result applies to the graph induced by $y$ and hence, by \Claim{equiv}, also to $x$:
\begin{equation}
\EquationName{grow}
\card{ \setst{ U }{ \emptyset \neq U \subsetneq V ,~ x(\delta^+(U)) \leq \alpha }}
~\leq~ n^{2 \alpha}.
\end{equation}

To round $x$, we must first scale it so that its minimum cut value is large.
Let $G$ be the directed, weighted, multigraph obtained from $x$ by taking
$K \defeq 100 \ln n$ parallel copies of each edge, each of the same weight as in $x$.
Let $c_i$ be the value of the $i\th$ cut, ordered such that $K \leq c_1 \leq c_2 \leq \cdots$.
We will construct a directed multigraph $H$ by taking each edge of $G$ with probability proportional
to its weight.
The expected number of edges chosen by $H$ in the $i\th$ cut is $c_i$.
Let $p_i$ be the probability that the actual number of edges chosen in the $i\th$ cut
diverges from its expectation by more than an $\epsilon$ fraction.
By a Chernoff bound, $p_i \leq 2 e^{-\epsilon^2 c_i / 3}$.

We will ensure that no cut diverges significantly from its expectation by
choosing $\epsilon$ appropriately and applying a union bound.
Define $\epsilon = \sqrt{1/10}$.
Since $c_i \geq K = 100 \ln n$, we have $p_i \leq n^{-3}$ for all $i$.
For the small cuts, we use the bound
\begin{equation}
\EquationName{small}
\sum_{i=1}^{2n^2} p_i ~=~ O(1/n).
\end{equation}
For the large cuts, we use a different bound.
\Equation{grow} implies that $c_{n^{2\alpha}} \geq \alpha K$.
Letting $i = n^{2\alpha}$, we have $c_i \geq K \ln i/(2 \ln n)$,
and hence $p_i \leq i^{-3/2}$.
Thus
\begin{equation}
\EquationName{big}
\sum_{i>n^2} p_i
~\leq~ \sum_{i>n^2} i^{-3/2}
~<~ \int_{n^2}^{\infty} x^{-3/2} \,dx
~=~ O(1/n).
\end{equation}
Combining \Equation{small} and \Equation{big} shows that
with probability $1-O(1/n)$, no cut in $H$ diverges from its expectation by
more than an $\epsilon$ fraction.
The expected cost of $H$ is $K \cdot c \transpose x$,
so a Chernoff bound again implies that the cost of $H$ is $O(\log n) \cdot c \transpose x$
with high probability.

Let $z \in \bZ_+^{V \times V}$ be the vector giving the total weight of the edges in the multigraph $H$.
Assuming that no cut in $H$ diverges significantly from its expectation,
we have
\begin{equation}
\EquationName{balanced}
\frac{z(\delta^+(U))}{z(\delta^-(U))}
~\leq~ \frac{1+\epsilon}{1-\epsilon}
~\leq~ 2
\qquad\forall\,\emptyset \neq U \subsetneq V.
\end{equation}
The last inequality follows because $\epsilon < 1/3$. Thus  $z$ is a nearly-balanced graph with high probability.

\subsection{Patching Up}
We now make $z$ Eulerian by ``patching it up'' with another graph $w$.
That is, we seek another vector $w \in \bZ_+^{V \times V}$ such that
$z+w$ is connected and Eulerian ---
an integral, feasible solution to the Held-Karp relaxation of ATSP. Indeed, we show that
a subgraph of $z$ can be used to patch up $z$.

Consider the transshipment problem on $V$ where each vertex $v$ has demand
$b(v) := z(\delta^+(v))-z(\delta^-(v))$.
Hoffman's circulation theorem \cite[Corollary 11.2f]{Schrijver}
implies that there exists
a subgraph of $z$ giving a feasible, integral transshipment for these demands iff
the capacity of each cut is at least its demand:
$$
z(\delta^-(U))
~\geq~ \sum_{v \in U} b(v)
~=~ z(\delta^+(U))-z(\delta^-(U)).
$$
This inequality is implied by \Equation{balanced}, so the desired transshipment $w$ exists,
and its cost is at most $c \transpose z$.
Thus $z+w$ gives a connected, Eulerian graph of cost at most $2 \, c \transpose z$,
which is $O(\log n) \cdot c \transpose x$, as argued above.
By shortcutting, we obtain a tour of no worse cost.
If $x$ is an optimum solution of the linear program
then the resulting tour is at most a factor $O(\log n)$ larger than the optimum tour.
Consequently, the integrality gap of this linear program is at most $O(\log n)$.

\section{Tight Example}

We now show that the analysis of the algorithm given above is tight to within constant factors ---
we give an example where we must choose $K$ to be
$\Omega(\log n)$ in the first step of the algorithm.
This condition is necessary not only
to ensure the nearly-balanced condition but also to ensure that $H$ is connected. 

Consider any extreme point $x$ such that $x_a<\frac{2}{3}$ for every arc $a\in A$ and let $E$ be the
support of $x$. Such extreme points exist of arbitrarily large size~\cite{CGK}.
Using the fact that $\card{E}<3n-2$ where $n=\card{V}$, we obtain that there in
an independent set of vertices $V_1$ of size at least $\frac{n}{6}$.
Since $x_a<\frac{2}{3}$ for each $a\in A$ and $x(\delta^-(v))+x(\delta^+(v))=2$, the probability
that $v$ is a not an isolated vertex in $H$ is at most $1-\frac{1}{27^K}$ for each $v\in V_1$.
Since $V_1$ is an independent set, these events are independent.
Hence, the probability that none of the vertices in $V_1$ is an isolated vertex in $H$ is at most
$(1-\frac{1}{27^K})^{\frac{n}{6}}$.
In order for $H$ to be connected with constant probability, we must take $K=\Omega(\log n)$.


\end{document}